\documentclass[longauth,traditabstract]{aa}
\usepackage{aas_macros}
\usepackage{natbib}
\usepackage{graphicx}
\usepackage{amssymb}
\newcommand{\degrees}{\ensuremath{^\circ}}
\begin{document}
\title{LBCS: the LOFAR Long-Baseline Calibrator Survey }
\titlerunning{LBCS}
\author{Neal Jackson$^1$, Amitpal Tagore$^1$, Adam Deller$^2$, Javier Mold\'on$^{2,1}$, Eskil Varenius$^3$, Leah Morabito$^4$, Olaf Wucknitz$^5$, Tobia Carozzi$^3$, John Conway$^3$, Alexander Drabent$^6$, Anna Kapinska$^{7,8,9}$,  Emanuela Orr\`u$^2$, Michiel Brentjens$^2$, Richard Blaauw$^2$, Geert Kuper$^2$, Jurjen Sluman$^2$, Jorrit Schaap$^2$, Nico Vermaas$^2$, Marco Iacobelli$^2$, Luciano Cerrigone$^2$, Aleksandar Shulevski$^2$, Sander ter Veen$^2$, Richard Fallows$^2$, Roberto Pizzo$^2$, Mike Sipior$^2$, 
J.~Anderson\inst{10}\and 
I.~M.~Avruch\inst{11,12}\and 
M.~E.~Bell\inst{13}\and 
I.~van Bemmel\inst{14}\and 
M.~J.~Bentum\inst{2,15}\and 
P.~Best\inst{16}\and 
A.~Bonafede\inst{17}\and 
F.~Breitling\inst{18}\and 
J.~W.~Broderick\inst{2}\and 
W.~N.~Brouw\inst{2,12}\and 
M.~Br\"uggen\inst{17}\and 
B.~Ciardi\inst{19}\and 
A.~Corstanje\inst{20}\and 
F.~de Gasperin\inst{4}\and 
E.~de Geus\inst{2,21}\and 
J.~Eisl\"offel\inst{5}\and 
D.~Engels\inst{22}\and 
H.~Falcke\inst{20,2}\and 
M.~A.~Garrett\inst{2,4}\and 
J.~M.~Grie\ss{}meier\inst{23,24}\and 
A.~W.~Gunst\inst{2}\and 
M.~P.~van Haarlem\inst{2}\and 
G.~Heald\inst{13,2,12}\and 
M.~Hoeft\inst{6}\and 
J.~H\"orandel\inst{20}\and 
A.~Horneffer\inst{5}\and 
H.~Intema\inst{4,25}\and 
E.~Juette\inst{26}\and 
M.~Kuniyoshi\inst{27}\and 
J.~van Leeuwen\inst{2,28}\and 
G.~M.~Loose\inst{2}\and 
P.~Maat\inst{2}\and 
R. McFadden\inst{2}\and 
D.~McKay-Bukowski\inst{29,30}\and 
J.~P.~McKean\inst{2,12}\and 
D.~D.~Mulcahy\inst{1}\and 
H.~Munk\inst{31,2}\and 
M.~Pandey-Pommier\inst{32}\and 
A.~G.~Polatidis\inst{2}\and 
W.~Reich\inst{5}\and 
H.~J.~A.~R\"ottgering\inst{4}\and 
A.~ Rowlinson\inst{2,28}\and 
A.~M.~M.~Scaife\inst{1}\and 
D.~J.~Schwarz\inst{33}\and 
M.~Steinmetz\inst{18}\and 
J.~Swinbank\inst{34}\and 
S.~Thoudam\inst{20}\and 
M.~C.~Toribio\inst{4,2}\and 
R.~Vermeulen\inst{2}\and 
C. Vocks\inst{18}\and 
R.~J.~van Weeren\inst{35}\and 
M.~W.~Wise\inst{2,28}\and 
S.~Yatawatta\inst{2}\and 
P.~Zarka\inst{36}}
\authorrunning{Jackson et al.}
\institute{$^1$University of Manchester, School of Physics and Astronomy, Jodrell Bank Centre for Astrophysics, Oxford Road, Manchester M13 9PL, UK\\
$^2$ASTRON, Netherlands Institute for Radio Astronomy, Postbus 2, 7990 AA Dwingeloo, Netherlands\\
$^3$Department of Earth \& Space Sciences, Chalmers University of Technology, Onsala Space Observatory, 43992 Onsala, Sweden\\
$^4$Sterrewacht Leiden, University of Leiden, 2300RA Leiden, Netherlands\\
$^5$Max-Planck Institut f\"ur Radioastronomie, Auf dem H\"ugel 69, 53121 Bonn, Germany\\
$^6$Landessternwarte Tautenburg, Sternwarte 5, 07778 Tautenburg, Germany\\
$^7$University of Portsmouth, Winston Churchill Ave, Portsmouth PO1 2UP, United Kingdom\\
$^8$International Centre for Radio Astronomy Research (ICRAR), University of Western Australia, Crawley WA 6009, Australia\\
$^9$CAASTRO, Arc Centre of Excellence for All-Sky Astrophysics,  Australia\\
$^{10}$Helmholtz-Zentrum Potsdam, DeutschesGeoForschungsZentrum GFZ, Department 1: Geodesy and Remote Sensing, Telegrafenberg, A17, 14473 Potsdam, Germany \\
$^{11}$SRON Netherlands Insitute for Space Research, PO Box 800, 9700 AV Groningen, The Netherlands \\
$^{12}$Kapteyn Astronomical Institute, PO Box 800, 9700 AV Groningen, The Netherlands \\
$^{13}$CSIRO Astronomy and Space Science, 26 Dick Perry Avenue, Kensington, WA 6151, Australia  \\
$^{14}$Joint Institute for VLBI in Europe, Dwingeloo, Postbus 2, 7990 AA The Netherlands \\
$^{15}$University of Twente, The Netherlands \\
$^{16}$Institute for Astronomy, University of Edinburgh, Royal Observatory of Edinburgh, Blackford Hill, Edinburgh EH9 3HJ, UK \\
$^{17}$University of Hamburg, Gojenbergsweg 112, 21029 Hamburg, Germany \\
$^{18}$Leibniz-Institut f\"{u}r Astrophysik Potsdam (AIP), An der Sternwarte 16, 14482 Potsdam, Germany \\
$^{19}$Max Planck Institute for Astrophysics, Karl Schwarzschild Str. 1, 85741 Garching, Germany \\
$^{20}$Department of Astrophysics/IMAPP, Radboud University Nijmegen, P.O. Box 9010, 6500 GL Nijmegen, The Netherlands \\
$^{21}$SmarterVision BV, Oostersingel 5, 9401 JX Assen \\
$^{22}$Hamburger Sternwarte, Gojenbergsweg 112, D-21029 Hamburg \\
$^{23}$LPC2E - Universite d'Orleans/CNRS \\
$^{24}$Station de Radioastronomie de Nancay, Observatoire de Paris - CNRS/INSU, USR 704 - Univ. Orleans, OSUC , route de Souesmes, 18330 Nancay, France \\
$^{25}$National Radio Astronomy Observatory, 1003 Lopezville Road, Socorro, NM 87801-0387, USA \\
$^{26}$Astronomisches Institut der Ruhr-Universit\"{a}t Bochum, Universitaetsstrasse 150, 44780 Bochum, Germany \\
$^{27}$National Astronomical Observatory of Japan, Japan \\
$^{28}$Anton Pannekoek Institute for Astronomy, University of Amsterdam, Science Park 904, 1098 XH Amsterdam, The Netherlands \\
$^{29}$Sodankyl\"{a} Geophysical Observatory, University of Oulu, T\"{a}htel\"{a}ntie 62, 99600 Sodankyl\"{a}, Finland \\
$^{30}$STFC Rutherford Appleton Laboratory,  Harwell Science and Innovation Campus,  Didcot  OX11 0QX, UK \\
$^{31}$Radboud University Radio Lab, Nijmegen, P.O. Box 9010, 6500 GL Nijmegen, The Netherlands \\
$^{32}$Centre de Recherche Astrophysique de Lyon, Observatoire de Lyon, 9 av Charles Andr\'{e}, 69561 Saint Genis Laval Cedex, France \\
$^{33}$Fakult\"{a}t f\"{u}r Physik, Universit\"{a}t Bielefeld, Postfach 100131, D-33501, Bielefeld, Germany \\
$^{34}$Department of Astrophysical Sciences, Princeton University, Princeton, NJ 08544, USA \\
$^{35}$Harvard-Smithsonian Center for Astrophysics, 60 Garden Street, Cambridge, MA 02138, USA \\
$^{36}$LESIA \& USN, Observatoire de Paris, CNRS, PSL/SU/UPMC/UPD/SPC, Place J. Janssen, 92195 Meudon, France \\
}

\clearpage
\date{\today}
\abstract{
We outline LBCS (the LOFAR Long-Baseline Calibrator Survey), whose aim is to identify sources suitable for calibrating the highest-resolution observations made with the International LOFAR Telescope, which include baselines $>$1000 km. Suitable sources must contain significant correlated flux density ($\gtrsim 50 - 100$ mJy) at frequencies around 110--190~MHz on scales of a few hundred milliarcseconds. At least for the 200--300-km international baselines, we find around 1 suitable calibrator source per square degree over a large part of the northern sky, in agreement with previous work. This should allow a randomly selected target to be successfully phase calibrated on the international baselines in over 50\% of cases. Products of the survey include calibrator source lists and fringe-rate and delay maps of wide areas -- typically a few degrees -- around each source. The density of sources with significant correlated flux declines noticeably with baseline length over the range 200--600~km, with good calibrators on the longest baselines appearing only at the rate of 0.5 per square degree. Coherence times decrease from 1--3 minutes on 200-km baselines to about 1 minute on 600-km baselines, suggesting that ionospheric phase variations contain components with scales of a few hundred kilometres. The longest median coherence time, at just over 3 minutes, is seen on the DE609 baseline, which at 227~km is close to being the shortest. We see median coherence times of between 80 and 110 seconds on the four longest baselines (580--600~km), and about 2 minutes for the other baselines. The success of phase transfer from calibrator to target is shown to be influenced by distance, in a manner that suggests a coherence patch at 150-MHz of the order of 1 degree. Although source structures cannot be measured in these observations, we deduce that phase transfer is affected if the calibrator source structure is not known. We give suggestions for calibration strategies and choice of calibrator sources, and describe the access to the online catalogue and data products.}
\maketitle

\section{Introduction}

Calibration of effects due to the Earth's atmosphere is a crucial part of imaging with radio interferometers. At low radio frequencies, of a few hundred Megahertz (MHz) or less, the dispersive effect of the Earth's ionosphere is the principal propagation effect that corrupts the interferometric visibility data. It does this by imposing a rapidly-varying corrugation in the wavefront, that along a given line of sight can vary by a radian or more on a timescale of minutes. If this effect is not removed, each baseline of the interferometer will contain a phase term that varies randomly on a short timescale, and the phase coherence needed to form fringes of astronomical objects will be lost. To achieve this coherence, a calibrator source must be observed, whose structure is known, that is bright enough to allow the propagation effects to be solved for, and that is close enough to the target on the sky to be subject to (approximately) the same propagation effects as the target.

\begin{figure}
\includegraphics[width=8cm]{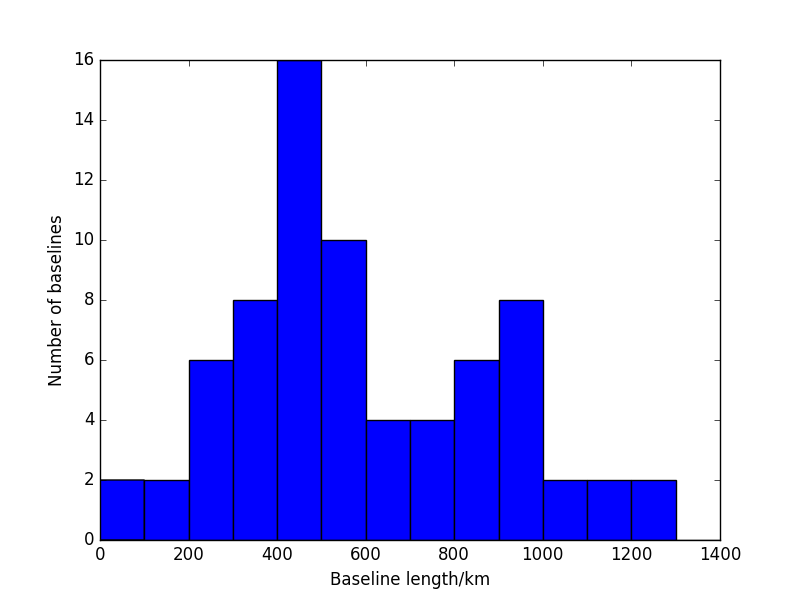}
\caption{Distribution of LOFAR international array baseline lengths.}
\label{baselines}
\end{figure}

The LOFAR telescope \citep{van-haarlem13a} consists of 40 stations within the Netherlands (24 ``core stations'' close to the centre of the array in Exloo and 16 ``remote stations'' further away), six stations in Germany (Unterweilenbach, Potsdam, Effelsberg, J\"ulich, Tautenburg, and Norderstedt), and one each in the UK, France and Sweden (respectively at Chilbolton, Nan\c{c}ay and Onsala). Three stations (\L{}azy, Baldy and Bor\'owiec) in Poland are also now coming into operation, and a further station at Birr in Ireland is to be constructed. Each station consists of two antenna arrays, a Low-Band Array (LBA) covering the wavelength range 30-90~MHz, and a High-Band Array (HBA) covering 110--240~MHz. The core stations may be combined by insertion of phase offsets into a single ``tied station'' with a much larger collecting area. The LOFAR baseline lengths between stations range from a few tens of metres in the centre of the array, to baselines of, for example, 1300~km between Nan\c{c}ay and Onsala and nearly 2000~km to the Polish stations. Angular resolutions of $\sim$250~mas are therefore possible at HBA frequencies. Fig.~\ref{baselines} gives a histogram of baseline lengths from the stations, excluding the Polish stations, which were not used in this phase of the LBCS observations. 

Calibration of LOFAR observations with the Dutch ($<$80-km) baselines is relatively straightforward, at least once an approximate model of the sky in the field of interest is available. Because the fields of view are large, typically a few degrees, there are many bright radio sources visible, providing an abundance of correlated flux on 5--10 arcsecond scales that can be used to calibrate the atmospheric phase effects on short timescales. The major challenge in this case is to provide calibration algorithms that can cope with the complexity of simultaneous estimation of atmospheric effects in many different directions at once. In international-baseline LOFAR imaging, the challenges are rather different. Here, the data are typically averaged in time and frequency so that the field of view is small; however, because LOFAR stations consist of phaseable antenna arrays, it is possible to arrange several beams on the sky and hence look in several different directions at once. In principle, it is possible to put one beam, with a fraction of the available bandwidth, on a calibrator source and the rest on the target. The phase solutions can then be made on the calibrator and transferred to the target before imaging. The whole procedure is similar to traditional VLBI in that phase solutions need to be transferred across space, and that single sources can be processed largely independent of each other; and VLBI now also allows data to be correlated in different spatial directions so that separate fields on target and calibrator can be maintained.

However, in order to perform this calibration, the phase calibrator source must contain a compact component that generates correlated signals even on the longest baselines, which for International LOFAR means spatial scales of $<1^{\prime\prime}$. At low frequencies, radio sources typically consist of optically-thin synchrotron emission in large-scale structures such as the radio lobes of extended radio galaxies, which are resolved out and effectively invisible to long baselines. Flat-spectrum sources, dominated by unresolved cores and which make good VLBI calibrators at GHz frequencies, tend to consist of self-absorbed synchrotron components that become optically thick and hence decrease in flux density at lower frequencies. The combination of these two effects make it difficult to find good low-frequency calibrators for long baselines. 

Some previous VLBI studies have been done at $\sim$100~MHz frequencies. In particular, Clark et al. (1975)\nocite{clark75} observed about 100 strong radio sources using baselines across the U.S., detecting nearly all sources on 50-km baselines, but observing that the correlated flux decreased, or became undetectable, in the vast majority of sources on $\sim$2500-km baselines. In addition, regular observations at 100--200~MHz frequencies were done with long-baseline arrays such as MERLIN in the subsequent decades (e.g. Leahy, Muxlow \& Stephens 1989)\nocite{leahy89}. Observations were also conducted using VLBI at slightly higher frequencies, for applications ranging from extensive observations of individual objects to wider-field surveys (e.g. Ananthakrishnan et al. 1989, Lenc et al. 2008)\nocite{ananthakrishnan89,lenc08}. The LOFAR international baselines follow in a long tradition of low-frequency work with long baselines, but with the potential to conduct large-scale surveys of the whole sky. Previous work, however, gives us an indication of the likely behaviour of correlated flux as a function of baseline length (which ranges from 200~km to over 1000~km) in observations with the LOFAR array. 

In principle an observed field can be searched for calibrators by averaging a large dataset repeatedly around different positions, corresponding to possible calibrator sources, but this is time-consuming and awkward. Once the Polish stations join the array, it will be difficult to carry out such procedures because the usable field of view provided by the pre-averaged LOFAR time and frequency resolution will shrink, making it unlikely that a randomly selected pointing will contain a good calibrator. It is therefore necessary to find the calibrators, in order that any observation can be calibrated using a separate beam on a source known to have compact structure.

A pilot calibrator survey was carried out by \citet{moldon15a}. They studied a small area (approximately 100 square degrees) of sky, using four-minute exposures with 30 beams formed simultaneously on 30 different sources chosen from the 74-MHz VLA Low-frequency Sky Survey \citep[VLSS;][]{cohen07a,lane14a} and the 327-MHz Westerbork Northern Sky Survey \citep[WENSS][]{rengelink97a}. Sources were processed using standard LOFAR long-baseline analysis routines, and fringe-fitted to determine the baselines on which source structure was visible. The density of usable calibrators was estimated to be about 1 source per square degree, resulting in about 100 good calibrators being found over a small area of sky. As will be argued later (Section 4) this is sufficient for good phase calibration of most target sources.

This paper reports on a programme which builds on the \citet{moldon15a} work, with the intention of extending the search for LOFAR long-baseline calibrators to the whole northern sky. In Section 2 we describe the survey and the selection process. In Section 3 we outline the methods used to process the data and discuss the survey products. In Section 4 we discuss the implications of the LBCS results for phase and delay coherence on long LOFAR baselines.  In Section 5 we recommend calibration procedures for future observations with the long baselines of LOFAR, summarise the work and present the conclusions. We present the form of the survey data products in Appendix A.

\section{Survey selection and observations}

\subsection{Selection}

Our source selection in the region north of 30$^{\circ}$N uses three surveys. The first two surveys are the 74-MHz VLSS \citep[where we use the improved ``redux'' processing;][]{lane14a} and 327-MHz WENSS \citep{rengelink97a}. The third survey is the Multi-frequency Snapshot Sky Survey \nocite{heald15} (MSSS: Heald et al. 2015, Heald et al. 2016, in preparation) which was conducted with the LOFAR HBA, but with lower resolution, at 120-160~MHz. Where this was available, it was used to provide additional information for spectral index selection.  Recently, the first alternative data release for the TIFR GMRT Sky Survey \citep[ADR1 of the TGSS;][]{intema16a} was made; like MSSS, this provides information at $\sim$150 MHz.  ADR1 of TGSS was not available at the time of scheduling and data reduction for the information presented in this paper; we plan to make use of it in the future.

The basic LBCS sample above 30$^{\circ}$N consists of all WENSS sources which are identified as single in the WENSS catalogue, and which lie within 30$^{\prime\prime}$ of a source in either MSSS (in the region 7-17h and declination 30$^{\circ}$-60$^{\circ}$) or VLSS. There are 30686 such sources. In order to prioritise the observations, we note that \citet{moldon15a} investigated the fraction of good calibrators as a function of a number of source properties. They found that the main predictors for compact structure were a high total WENSS flux density, and a flat low-frequency spectrum; neither finding was unexpected, since compact and total flux density would be expected to correlate, and flat spectra in general indicate the presence of small, self-absorbed radio core structure. However, flat spectra at high (GHz) frequencies were not good predictors of compact structure. Again this is easily understandable, in that the classic GHz core-dominated sources have lower-frequency spectral turnovers due to the onset of synchrotron self-absorption, and their flux density decreases (in theory as $\nu^{5/2}$, although in practice less steeply) towards lower frequency. We nevertheless include GHz VLBI calibrators, in the form of the NRAO VLBA calibrator list, in the observing schedule. In principle we could also add pulsars, which are currently selected against due to their steep spectra, although in practice they were omitted in order to keep the survey selection criteria as simple as possible. In the TGSS ADR1 \cite{intema16a} there are about 90 pulsars with flux densities $>$100mJy at 150~MHz, which would be suitable as LBCS calibrators, in 37000 square degrees. We will add these to a future version of the catalogue, but for the moment remark that these are also suitable as long-baseline calibrator sources.

Within the northern sample, we define a goodness parameter as

\begin{equation}
g = 2.0+\log_{10}S+2.0\alpha
\end{equation}

\noindent where $S$ is the WENSS flux density, in Jy, and $\alpha$ is the 74~MHz to 327~MHz spectral index\footnote{We define spectral index $\alpha$ as $S(\nu)=\nu^{\alpha}$ throughout, where $S$ is the flux density at a frequency $\nu$.}. This parameter is motivated by Fig. 7 of Mold\'on et al. (2015), in which a line of this gradient is the most efficient way of separating good calibrators from other sources, and $g>0$ defines the region of $S$ vs. $\alpha$ in which nearly all good calibrators are found. We intend to observe all sources within the sample for which $g>0$, together with other sources as necessary in order to obtain as dense a network of calibrators as possible. Due to the available observing time, the cut is currently made at $g=0.096$.

Below 30$^{\circ}$N WENSS is not available, and we therefore use coincidences (within 30$^{\prime\prime}$ in position) between VLSS and the 1.4-GHz NRAO VLA Sky Survey \citep[NVSS;][]{condon98a}. In this case, we demand that the NVSS source be unresolved, to give the maximum chance that a significant fraction of the flux density is in compact structure. This is by no means guaranteed, however, since the resolution of NVSS is 45$^{\prime\prime}$, a factor of 100 greater than the scale on which we are looking for compact structure. It is therefore likely that the LBCS search observations already made will be less efficient in the southern part of the sample.  Future observations to complete LBCS will be able to make use of TGSS or MSSS, which will compensate for the lack of WENSS information.  If necessary, we will conduct a second campaign to improve the density of the calibrator grid south of 30$^{\circ}$N using MSSS or TGSS data to identify additional candidate sources.

\subsection{Observations}

Observations were conducted on a number of occasions during 2014--2015, which are listed in Table \ref{observations}. Observations were conducted with the High Band Array (HBA) in the HBA DUAL-INNER mode. Each observation consisted of a six-minute cycle including one minute on a calibrator and three minutes of observation of targets, the remainder being setup time in between observations. The calibrator, normally 3C~196 or 3C~295, was used for phasing of the core stations (Section 3). The target observations consisted of 30 separate beams, pointed at different sources in the sample. When observing in 8-bit mode, LOFAR has a total available bandwidth of 96 MHz, divided into 488 sub-bands of width 0.195 MHz.  In our observing setup, each beam consisted of 16 such sub-bands spanning the frequency range 139-142 MHz, with each sub-band divided into 64 channels. The observing parameters are summarised in Table~\ref{parameters}.

Because a separate station beam is formed on each source, we enjoyed considerable freedom in selecting sources across the sky for any individual group of 30 beams.  However, the analogue tile beam-forming used for the LOFAR HBA system does impose a ``tile'' beam response with approximate full width at half maximum (FWHM) of 30\degrees.  In practice, the beams were allocated such that the sources in any group of 30 observations lay within a radius of 5\degrees\ of the centre of the tile beam, well within the tile beam half-power point. The calibrator observations were not simultaneous with the target observations, because the calibrators usually lay outside the tile beam.

In the two sets of observations taken in March and April 2015, six hours' worth of observations were repeated to assess the effect of different observing conditions, principally different ionospheric conditions but also to check for any variations of data quality at different times.

\begin{table}
\begin{tabular}{lcl} \hline
&&  \\
Observation &  No. of &  Stations  \\
date   &  sources &\\     
&&\\  \hline && \\
2014 Dec 18 & 210 & Not DE604,UK608,DE609  \\
2015 Mar 8-14 &  1437 & All stations \\ 
2015 Apr 2-7 &  6478 & All stations  \\
2015 Sep 28-Oct 2 & 5616 & All stations \\
2015 Nov 12-15 & 1184 & All stations  \\
&& \\ \hline \end{tabular}
\caption{Observation log. All stations (DE601 = Effelsberg, DE602 = Unterweilenbach, DE603 = Tautenburg, DE604 = Potsdam, DE605 = J\"ulich, FR606 = Nan\c{c}ay, SE607 = Onsala, UK608 = Chilbolton, DE609 = Norderstedt) were used unless otherwise stated.}
\label{observations}
\end{table}

\begin{table}
\begin{tabular}{lc} \hline
&\\
Parameter & Value \\
&\\ \hline
&\\
Array & LOFAR HBA \\
Configuration & DUAL\_INNER \\
Frequency & 139-142 MHz \\
Bandwidth per source & 3 MHz \\
Integration time per source & 3 min \\
Sources per observation & 30 \\ \hline
\end{tabular}
\caption{Observational parameters}
\label{parameters}
\end{table}

\section{Data processing and survey products}
\subsection{Data pre-processing}

The long-baseline pre-processing pipeline, developed by the LOFAR Long-Baseline Working Group and implemented by the Radio Observatory, was used to pre-process the data. This followed the procedure used by \citet{moldon15a}. Briefly, station gains were first solved for the Dutch stations using the BlackBoard Selfcalibration software \nocite{pandey09} ({\sc bbs}, Pandey et al. 2009) using a 1 minute scan on a ``primary'' calibrator (a primary calibrator scan was made in between every 3-minute target scan), before the core stations were combined to form a tied station, denoted ST001. In principle this should give a sensitivity equal to the combined stations forming the tied station, assuming uncorrelated noise and perfect phase calibration. In practice, the sensitivity achieved on baselines to ST001 was typically equivalent to about 3-5 core stations; since international stations are four times larger than core stations, this gives us in effect an extra international-sized station, while reducing the overall size of the dataset compared to using core stations separately. In some datasets we experimented with forming a second tied station, ST002, using the stations in the island (known as the {\it superterp}) in the centre of the core (stations CS002-CS007). Baselines to this station have a reduced sensitivity compared to ST001, but a wider field of view. The ``primary'' calibrator was a bright, arcsecond-scale source such as (depending on hour angle) 3C48, 3C196, 3C295 or 3C380. Calibration solutions were performed and applied to the data using the {\sc bbs} calibration routine, and the station addition was performed using the New Default Pre-Processing Pipeline ({\sc ndppp}). A priori station beam models were applied in {\sc bbs} before solutions were derived.

The corrected data, with the addition of the tied station(s), were then converted to circular polarization as described by \citet{varenius15a} and \citet{moldon15a}. This process means that Faraday rotation effects appear as a L-R phase difference, rather than shifting power from the parallel to the cross-hand linear polarisation products. The datafiles were then converted to FITS format, and all baselines to core stations, together with those to remote stations $<$20~km from the core, were removed to save disk space. The compressed dataset, after this removal, contains only the tied station, the remaining Dutch remote stations and the international stations. Finally, the dataset, averaged to 2 second time resolution and 4 channels per subband, was combined into a single intermediate frequency (IF) band with 64 channels each of 48.9~kHz width at 140~MHz (4 channels per subband $\times$ 16 subbands per source). The time and frequency resolutions of this averaged dataset correspond to unsmeared fields of view of about 1$^{\circ}$ and 0.5$^{\circ}$, respectively, beyond which fluxes of sources away from the observed source will be significantly reduced. The resulting datasets, comprising about 70~Mbytes per source, were read into the NRAO Astronomical Image Processing System, ({\sc aips})\footnote{{\sc aips} is distributed by the US National Radio Astronomy Observatory, {\tt www.nrao.edu}.}, for further analysis.

\subsection{Fringe fitting}

Each pre-processed dataset consists of about 90 visibilities per baseline, each with 64 frequency channels. If the source consists of an unresolved point at the phase centre, the data consist of a uniform amplitude and zero phase on all visibilities. However, extra delays may impose a slope in the phase of the visibilities as a function of frequency, and varying phases impose a gradient of phase with time. Delays may be due to clock offsets between stations, which give a constant time delay as a function of frequency, or to the ionosphere, in which case the delay increases with decreasing frequency. Ionospheric phase variations are clearly seen in all datasets, and can give phase gradients of a radian per minute, or more. Finally, an offset of the source from the phase centre can result in a gradient in both phase and delay, depending on the geometry of the baseline.

\begin{figure*}
\begin{tabular}{cc}
\includegraphics[width=9cm]{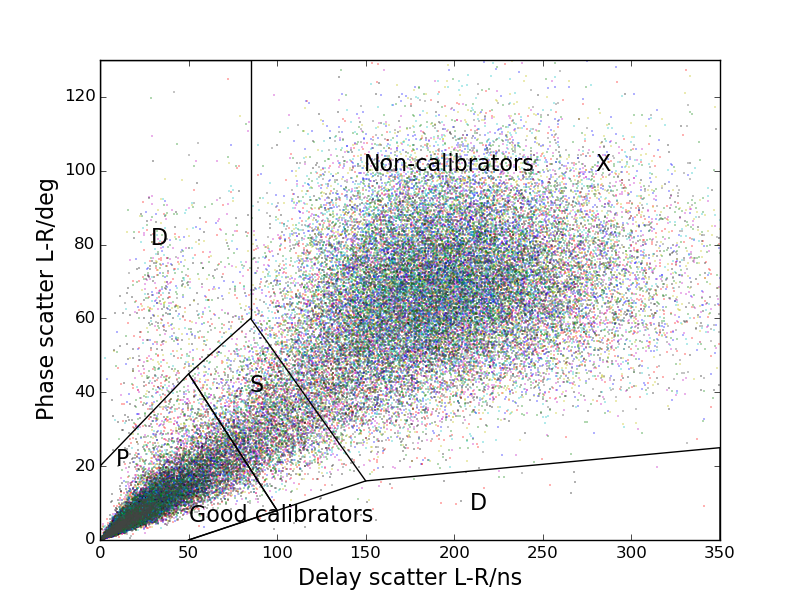}&\includegraphics[width=9cm]{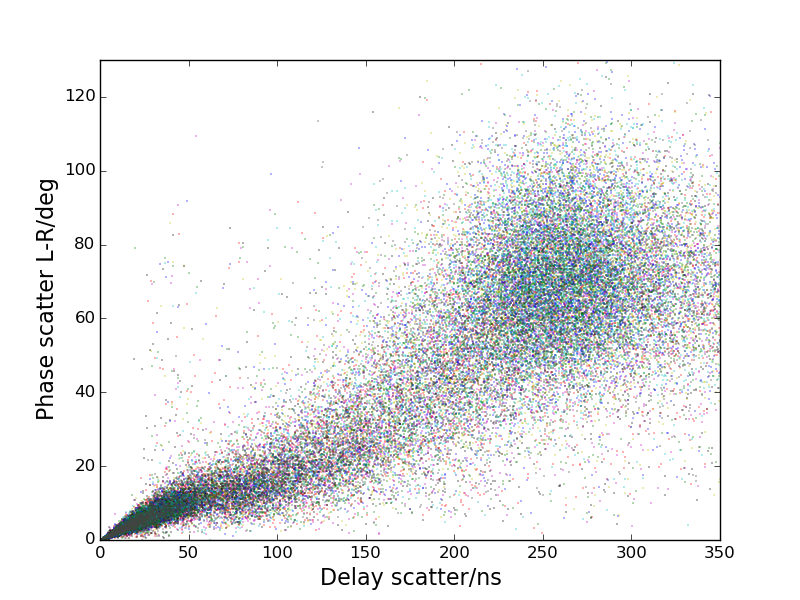}\\
\end{tabular}
\caption{Measures of strength of compact structure on baselines to ST001 from individual international stations, plotted with different colour points. The ordinate on each axis is the scatter in difference in phase solutions between L and R polarizations, which is independent of the atmospheric phase rotation and depends only on the signal-to-noise of structure on that baseline. On the left, the abscissa is the scatter in difference in delay solution between L and R, and on the right it is the absolute delay scatter on the L polarization. The diagram on the left shows the regions in which we consider calibrators to be good (P), marginal (S), or not enough correlated flux for calibration (X). Areas marked D are usually those with some problem with the data.}
\label{goodness}
\end{figure*}

The {\sc aips} program {\sc fring} was used to find the global fringe solution for each dataset. This program forms solutions by comparing the data with a model, which was initially taken as a point source in each case. The program solves for delay, rate (first derivative of the phase in the time direction) and phase, and also gives an indication of the  signal-to-noise of each solution. Rates can occur if, for example, the clocks drift with time. We set a delay search window of 500~ns, and a rate window of 5~mHz. We use a 6-second solution interval, enabling us to track non-linear variations of the visibility phase with time on sub-minute timescales if present. A source with strong compact structure will give a solution for both delay and phase with low scatter for any particular station, and will fail to give solutions, or give noisy solutions, if little flux density is seen on baselines to that station. An alternative method of assessing the level of structure is to take, for each station, the data on the baseline between it and the tied station, and Fourier transform it. The resulting 2-D image should contain a single bright point source if the source is compact on the spatial scale corresponding to that baseline. We discuss this further below.

For our measure of phase scatter, we consider the phase solutions on each baseline. The solution should be a smooth function of time, corresponding to atmospheric phase variations. We take this function, unwinding $2\pi$ phase differences between points where necessary, and compare the variation on the L polarization with that on the R polarization. Subtracting these two should give a constant in the limit of high signal-to-noise, which is not zero due to Faraday rotation combined with the conversion from linear to circular polarization. We take the scatter of this difference as the basic goodness statistic in phase (Fig.~\ref{goodness}). A similar procedure gives a goodness statistic in delay. These two quantities are plotted in Fig.~\ref{goodness} (left panel), and are used throughout the rest of the paper, and in the LBCS database, to define the quality of the calibrator for any of the baselines to the phased core station. It is clear that a set of high signal-to-noise solutions, corresponding to low phase and delay scatter, is seen for a population of source--baseline pairs labelled ``P'' on the diagram, and sources on particular baselines which do not have significant correlated flux density appear in a separate region (``X'') of the diagram, characterised by high delay and phase scatter corresponding to essentially random visibilities. In between is an area of sources ``S'' which are weakly detected. The remaining parts of the diagram are dominated by a small number of sources ``D'' where there are identified problems with the data, or where a neighbouring source within the field is confusing the delay and phase fits. Users should consult the supplementary information in such cases. It is very common for sources to appear as ``P'' on short baselines such as those between ST001 and DE601, DE605 and DE609 which are relatively close to the Dutch border. Considerably fewer sources give clear detections on the more remote stations, particularly UK608 and SE607. This already indicates that many sources become resolved at HBA frequencies at resolutions of about 0\farcs5. We do not assign a statistic to each source overall; users should determine appropriate calibrators based on the requirements corresponding to the range of baselines in any particular observation.

We note that \cite{moldon15a} used a slightly different statistic, related to the overall L-R delay difference with a long solution interval. Our delay statistic is similar to this, but has been chosen because it gives a rather better separation between the good and non-calibrator sources. We can also look at the correlated flux density of sources as a function of baseline length. Unsurprisingly, the furthest stations (Onsala, Nan\c{c}ay) give delay and phase solutions with higher scatter, since they  represent the baselines on which the source begins to be resolved out; this effect is discussed further below.

\begin{figure*}[ht]
\begin{tabular}{cc}
\includegraphics[width=9cm]{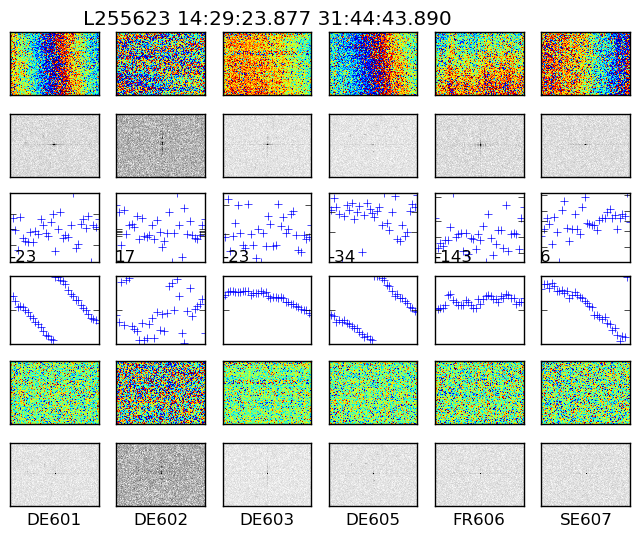}&\includegraphics[width=9cm]{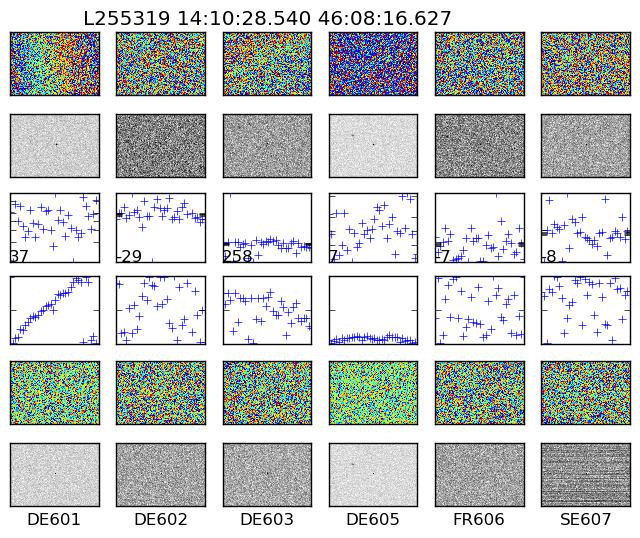}
\end{tabular}
\caption{Graphics from two typical detected sources, from the 2014 December 18 observations. On the left is the source J142923+314443, which is clearly detected on all stations. The columns in the diagram represent the individual telescopes (see Table~\ref{observations} for a list). The first row shows the phases in the data on the baseline between each station and the large tied station, plotted on a diagram whose abscissa represents increasing time and whose ordinate represents increasing frequency. The effect of ionospheric phase variation can be seen in the form of a phase change with time, particularly in DE601 and DE605 where one complete rotation is achieved in 4 minutes. The delay appears as a variation of phase with frequency, which here is quite moderate. The second row shows the Fourier transform of the images in the first row, clearly indicating the presence of a single compact source. DE602 was noisy during these observations, but nevertheless demonstrates that strong sources can be recovered. On the third and fourth rows, the delay and phase solutions are shown; the ticks on the $y$ axis represent ($-$50~ns, $-$20~ns, 0, +20~ns and +50~ns) from the median delay (which is indicated, in nanoseconds, by the number on each plot). The phase solutions in the fourth row are plotted in the range $-180^{\circ}$ to $+180^{\circ}$. The final two rows show the corrected data after the fringe fit and its Fourier transform; the residuals include a number of small phase jumps. On the right, the source J141028+460816 is shown. Here the source is clearly seen on the short baselines from the tied station in the Netherlands to DE601 and DE605, but the source is not suitable as a calibrator for the longer baselines.}
\label{example}
\end{figure*}

In Fig.~\ref{example} we show a montage of plots representing a well-detected source (which has compact structure on all baselines) and one which is only well detected to some stations. The well-detected source is characterised by low scatter in the delay solutions, a clear peak in the Fourier transform of the phase data, and clear structure in the phase solutions.  

\subsection{Fringe rate and delay mapping}

In addition to information on a single point source within each beam of each observation, we also obtain additional information on the field of each source in the form of a map of the sky constructed from fringe rates and delays in the data. This map contains other nearby sources, thus allowing us to check that compact sources are not being missed by our selection criteria; in the cases where we have performed visual inspection of the maps, the sources which appear do indeed appear in other observations. For this reason, and also because the maps are available as a database product, we briefly describe the process by which they are produced.

The interferometer phase equation (IPE) relates the phase corresponding to the geometrical path delay between any two telescopes, $\phi$, to the hour angle $H$ and declination $D$ of the source, and the hour angle $h$ and declination $d$ of the baseline (e.g. Rowson 1963)\nocite{rowson63}:

\begin{equation}
\phi = \frac{2\pi L}{\lambda}\left( \sin d \sin D + \cos d \cos D \cos (H-h) \right),
\end{equation}

\noindent where $L$ is the baseline length and $\lambda$ the observing wavelength. 

We can write the time and frequency derivatives of $\phi$ as functions of sky coordinates, expressed as offsets $\Delta H$ and $\Delta D$, as 

\large
\begin{equation}
\left( \begin{array}{c}
\partial\phi/\partial t \\
\partial\phi/\partial f \\
\end{array} \right)
=
\left( \begin{array}{cc}
\frac{1}{\cos D} \frac{\partial^2\phi}{\partial H \partial t} &
\frac{\partial^2\phi}{\partial D \partial t} \\
\frac{1}{\cos D} \frac{\partial^2\phi}{\partial H \partial f} &
\frac{\partial^2\phi}{\partial D \partial f} \\
\end{array} \right)
\left( \begin{array}{c}
\Delta H  \\ \Delta D \\
\end{array} \right),
\label{frd_equation}
\end{equation}

\normalsize

\noindent where the $\cos D$ terms account for the curved sky geometry. The Fourier transform of the visibility data $V(t,f)$ over a small interval in time and frequency therefore gives a two-dimensional image whose axes are related to $\partial\phi/\partial t$ and $\partial\phi/\partial f$, which can be transformed into a map of the sky using the coefficients of the matrix in Equation~\ref{frd_equation}. These coefficients are straightforward to calculate using the IPE, where the baseline hour angles and declinations are calculated from the $u$, $v$ and $w$ coordinates of the data. In practice, we have short observations and have therefore used the mean $u$, $v$, $w$ coordinates from each short stretch of data to produce maps of the sky for each baseline to ST001, where the longer baselines give generally lower signal-to-noise images with a smaller field of view, and the baselines to the closer German stations give fields of view of a few degrees. These separate images have been signal-to-noise weighted and combined to produce a composite image.

Unlike a conventional interferometer map, the resolution of a fringe-rate -- delay image is controlled in one direction by the overall bandwidth of the observations, and in the other by the length of the time over which the Fourier transform is taken. In practice, this means that the resolution in this direction is limited by the coherence time of the atmosphere. For the LBCS observations, the 3-MHz bandwidth and the 3-minute integration time limit the resolution on a 200-km baseline to 100$^{\prime\prime}$ in the delay direction and $160^{\prime\prime}\sec d$ in the fringe rate direction, where $d$ is the declination of the baseline (e.g. Peckham 1973\nocite{peckham73}). On the other hand, the field of view of the observations is controlled by the time and frequency resolution after averaging (which has been set at 2s and 4 channels/subband in order to maintain manageable data volumes in the raw observations where the core stations are present), and by the primary beam of the largest telescope. The bandwidth and time smearing function is complex, but implies an amplitude reduction of 30\% at 1 degree and frequency smearing of 40\% at 0.5 degrees for a 1000 km baseline.

\begin{figure}
\includegraphics[width=9cm]{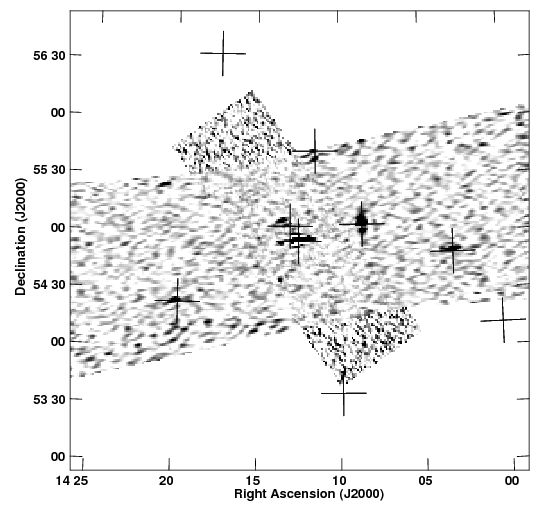}
\caption{Fringe-rate and delay map of a source; this is an unusually crowded field, and many other maps show only the targeted source. Each superposed parallelogram is the FRD map generated from an individual baseline to ST001, with the larger fields of view corresponding to the shorter baselines. The target source is clearly visible in the field, showing that it has compact structure, and in addition other sources are seen in the field which are also seen in the WENSS survey. All WENSS sources with flux density $>$400~mJy are shown by crosses.}
\label{frdmap}
\end{figure}

These images (Fig.~\ref{frdmap}) are subject to a number of distortions. In particular, delay offsets, and atmospheric phase gradients with time, will cause offsets of point sources in the frequency and time direction, respectively, and any differential delay across the frequency band (such as that produced by dispersive ionospheric delays) or non-linear phase variations will smear out the images. Moreover, the relatively small field of view of the ST001 station will reduce the amplitude of sources away from the field centre considerably. Nevertheless, these maps are very useful diagnostics, and give a good indication of the presence of sub-arcsecond, compact structure over the field. They can also be produced, as here, quickly and with very small quantities of data. In many cases, we also see other sources in the field, and can check their reality using other catalogues such as WENSS.

\subsection{Reproducibility}

As part of the observations, 6 hours were spent in repeating previous observations of some fields, in order to test the degree to which results were reproducible. Failure to reproduce results may be due to effects of the telescope system and data reduction, or, more likely, to variations in the atmospheric stability that would affect the ability of fringe fitting to detect the source. Nearly 1300 sources were re-observed, and separate observations are reported as separate rows in the database; each re-observation gives nine data points, one for each international station.

We generally find that observations are highly reproducible. Objects reported as well-detected (``P'') on a given baseline and on one of the observing epochs are recovered as well-detected sources in 80\% of the cases, and as marginally detected sources in a further 11\%. Similarly, objects found to be not detected (``X'') in one epoch are undetected in 85\% of other observations, and marginally detected in a further 8\%. Even in inconsistent cases, we typically find that only one or two telescopes return different results for a particular object, the other telescopes being consistent. We therefore conclude that the LBCS calibrators should be reliable, at least unless ionospheric conditions are very bad (in which case it is likely to be difficult in any case to transfer phase solutions from calibrators to the astronomical target).

\subsection{Correlation of source detection with low-frequency radio properties}

Mold\'on et al. (2015) found a clear tendency for sources detected on long baselines to be preferentially those sources that are bright at low frequencies, and those that have a relatively flat low-frequency spectral index. Neither conclusion is particularly surprising: sources with greater total flux density are likely to contain more correlated flux density, and sources with flatter radio spectral index are likely to contain more contributions to their flux density from physically small, and hence synchrotron self-absorbed, components. The LBCS selection is therefore biased towards bright sources with flat low-frequency spectra. Mold\'on et al. also found, however, that gigahertz-frequency spectral index was a relatively poor predictor of detectability on the long LOFAR baselines.

\begin{figure}
\includegraphics[width=9cm]{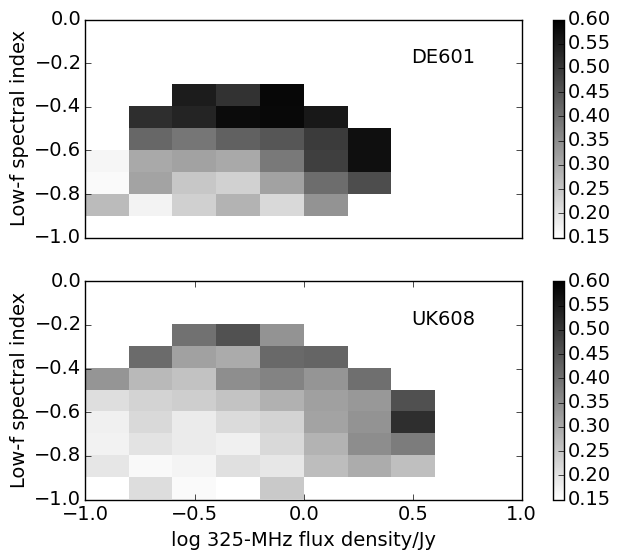}
\caption{Greyscale plot of the high signal-to-noise (category ``P'') detection fraction as a function of WENSS 325-MHz flux density and low-frequency spectral index. The latter quantity is calculated from WENSS together with the VLSS 74-MHz flux density and, where available the 6C 151-MHz flux density (Hales et al. 1993, and references therein). The upper plot represents the detections on a 266-km baseline, and the lower plot shows the detections on a 602-km baseline. Both these lengths are maximum projected baseline length, and in practice the projected baseline length will be less than this. Pixels with errors smaller than 0.1 are plotted. The line in each case represents $g=0$; in practice, the survey should be complete to $g=0.096$. A few objects with $g<0$ were observed, mostly because their MSSS flux densities indicated they might have a flat low-frequency spectrum.}
\label{detection}
\end{figure}

\nocite{hales93}

In Fig.~\ref{detection} we show the proportion of sources that are detected, as a function of flux density and low-frequency spectral index, for a typical short international baseline (the 266-km baseline between Effelsberg [DE601] and the Dutch core [ST001]) and a typical long baseline (the 602-km baseline between Chilbolton [UK608] and the Dutch core [ST001]). The previously noticed relationship between compact flux density and these two quantities is confirmed, with very low (10--20\%) detection rates for sources of 100~mJy and steep spectral indices, and about 50\% detections for flatter-spectrum sources of 1~Jy. We also clearly see the effect of baseline length. For example, a randomly selected 1-Jy source with a spectral index of $-$0.5 has a roughly 50\% chance of being a good calibrator on a 200-km baseline, but only 30\% at 600~km. It is likely that calibrators for 800--1000~km  baselines, such as those provided by the baselines to the new Polish stations, will need to be chosen carefully.

\subsection{Sky coverage and calibrator source density}

\begin{figure*}
\begin{tabular}{cc}
\includegraphics[width=9cm]{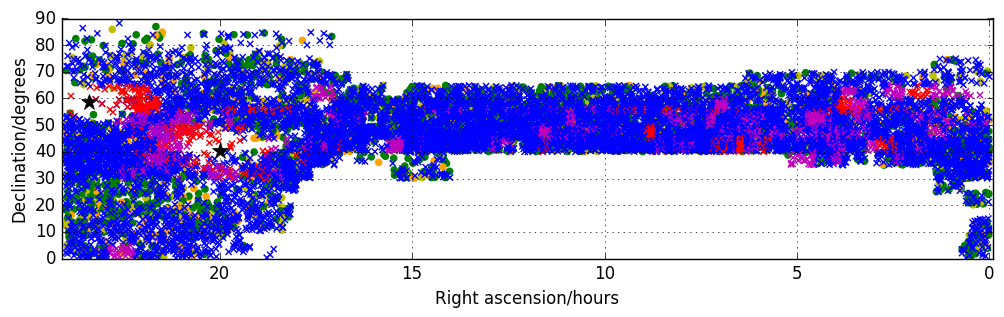}&
\includegraphics[width=9cm]{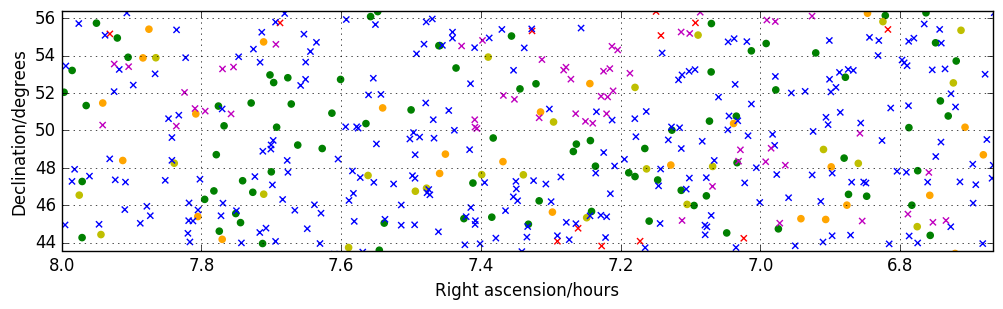}\\
\includegraphics[width=9cm]{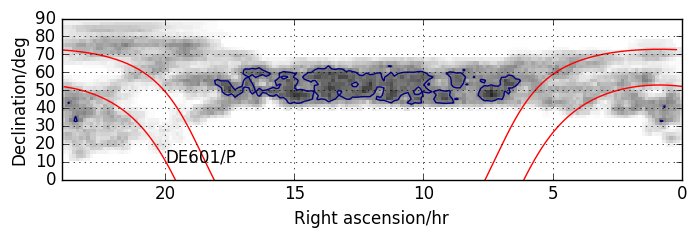}&
\includegraphics[width=9cm]{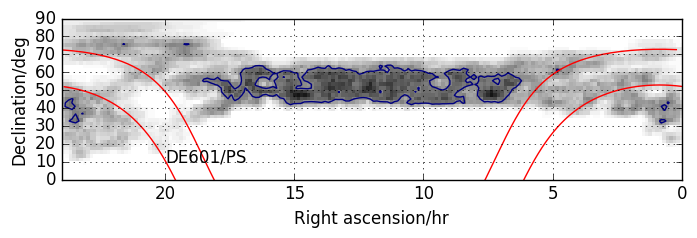}\\
\includegraphics[width=9cm]{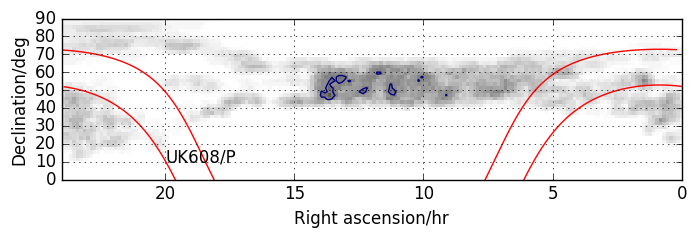}&
\includegraphics[width=9cm]{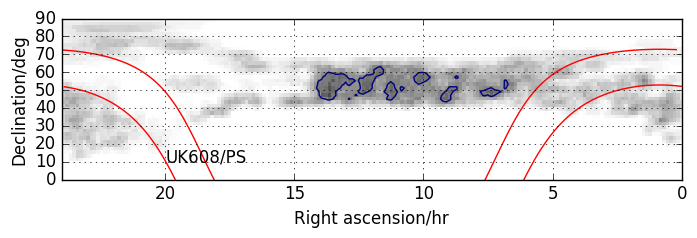}\\
\end{tabular}
\caption{Top left: current sky coverage of LBCS. Over subsequent observing seasons, the intention is to fill in gaps or areas where the observations were poor (see Section 3.6), as well as to extend the survey below 30$^{\circ}$N by selection from the MSSS survey. Top right: an expanded plot of a small region of sky. In both plots, green circles represent sources for which all, or all but one, observing stations gave P-class delay and phase solutions; greenish-yellow circles those for which all but one station gave P- or S-class solutions; and orange circles those for which all but two stations gave P- or S-class solutions. Blue crosses represent sources where the sources are likely not to be suitable as long-baseline calibrators. Purple and red crosses, respectively, represent the cases that are not detected on long baselines but the data is possibly, or definitely, faulty. Middle/lower: Contour plots of average density of calibrators, greyscaled from 0--2 per square degree and with a contour at 1 per square degree. Middle: calibrators with correlated flux density between Effelsberg (DE601) and the phased core (ST001). Lower: the same for Chilbolton (UK608). In each case, the left-hand plot represents the density of good (``P'') calibrators and the right-hand plot the density of good or marginal calibrators. The red lines in each plot represent Galactic latitude of $\pm 10^{\circ}$.}
\label{coverage}
\end{figure*}

Fig.~\ref{coverage} shows the current coverage of the LBCS survey. Currently, most of the sky above declination 40$^{\circ}$N has been covered, although there are regions where the data quality, as determined by the number of sources detected per observation, is not high. There are two significant holes in the coverage, where the source catalogue is sparse and most of the observed sources are not detected: one around RA=20$^{\rm h}$, Dec=40$^{\circ}$ and one around RA=23$^{\rm h}$30$^{\rm m}$, dec=60$^{\circ}$. These correspond to Cygnus A and Cassiopeia A, respectively, and in this area it is almost certain that the calibration of the core stations prior to forming the tied station (ST001 and/or ST002) has performed sub-optimally, yielding greatly reduced sensitivity.  We note that our calibration of the core stations did not include de-mixing \citep{van-der-tol07a} to mitigate the impact of bright out-of-beam sources such as Cygnus A and Cassiopeia A, due to the logistical challenges of de-mixing datasets that include the international stations, and so this poor performance is not unexpected.  In the near future, so-called ``smart'' de-mixing will become available, at which point a re-processing of these datasets would likely provide much improved results.

We have estimated the concentration of available calibrators on the sky as a function of position. This concentration has been calculated as the number of calibrators within 3$^{\circ}$ of a given position, and is effectively smoothed over a 3$^{\circ}$ radius. It therefore follows that this is an underestimate of the true concentration in regions close to the edge of the current coverage. Fig.~\ref{coverage} shows that, for regions away from the Galactic plane and Cas~A/Cyg~A, we easily obtain 1 good calibrator per square degree on the shorter (200-km) baselines. In the Galactic plane we are likely to be affected by scattering of radio waves in the Galactic disk (e.g. Cordes, Ananthakrishnan \& Dennison 1984) which can affect long-baseline and VLBI observations (e.g. Rickett \& Coles 1988; Rickett 1990). \nocite{cordes84,rickett88,rickett90} On the longer baselines, 1 good calibrator per square degree is obtained for only a small fraction of the area covered, although if we include the marginal (``S'') calibrators then this density is approached for about half of the covered area. This approximately agrees with the calibrator source densities found earlier by Mold\'on et al. (2015). This is an important number, because experience shows that in most cases it is desirable to have a calibrator within about 1--2\degrees\ of the target, depending on ionospheric conditions, for successful transfer of phases from phase calibrator to target.

The current sky coverage (February 2016) is about 7500 square degrees, and is complete between declinations +40$^{\circ}$ and +65$^{\circ}$, with patchy coverage outside these regions. Subsequent observing seasons in 2016 and 2017 will fill in first the northern sky above 30$^{\circ}$N, before observing the region between declinations 0$^{\circ}$ and +30$^{\circ}$. The database will be updated appropriately as the observations and data reduction progress, with the aim of finishing the project in late 2017.

\section{Phase and delay coherence on long LOFAR baselines}

\subsection{Atmospheric coherence time on international baselines}

The LBCS project gives us a large, homogeneous international-baseline dataset, covering a large fraction of the sky, at numerous epochs. It is therefore a useful dataset for investigating atmospheric effects on the data in a systematic way. In particular, the atmospheric coherence time, the typical delays and the isoplanatic patch are important. The first dictates the timescale over which phase calibration must be obtained, the second controls the ability to extrapolate phase calibration over the frequency band, and the third dictates the ability to extrapolate phase calibration from one source to another.

The coherence time can be calculated from the phase solution on each source that is sufficiently strong for the solution to be coherent -- in practice, this has been calculated on baselines to the tied station for which a ``P'' phase/delay solution is derived. Phase solutions with 6-second solution interval provided by {\sc fring} are used for this purpose. In each case, the phase solution is unwrapped to account for 2$\pi$ ambiguities, and a second-order polynomial is fitted to the results. This is done separately for the two parallel-hand polarization channels, and the best fit of these two to the data is chosen. This gives robustness against the occasional bad solution, and inspection by eye shows that this process yields results that reflect the phase variation well. The phase variation with time is then defined as the average absolute gradient of the polynomial representing the phase solution, and the coherence time is defined as the time needed for the phase change corresponding to this gradient to reach one radian. In principle we can measure arbitrarily long coherence times in this way, although we are not sensitive to any coherence time below $\sim$15 seconds, due to the 6-second solution interval.

\begin{figure}
\includegraphics[width=9cm]{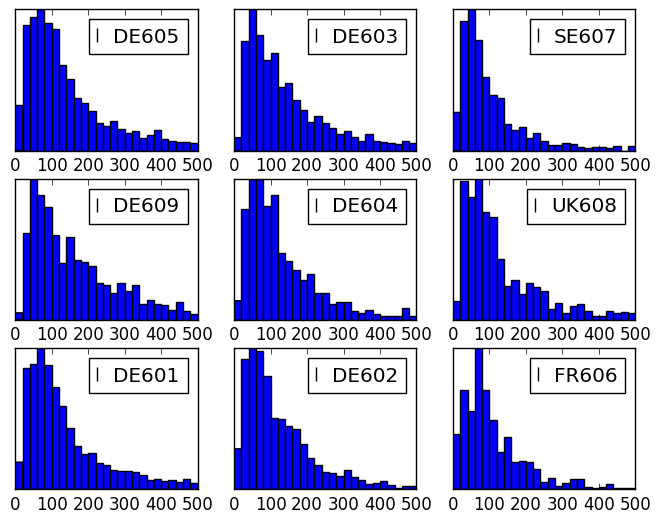}
\label{coherence}
\caption{Histograms of coherence time, in seconds, for each of the baselines from the international stations to the tied station at Exloo. Coherence times of 1--3 minutes are typical. Note that the short baselines (DE601, DE605, DE609) have noticeably longer average coherence times. The baseline lengths, from shortest to longest, are: DE605 (226~km), DE609 (227~km), DE601 (266~km), DE603 (396~km), DE604 (419~km), DE602 (581~km), SE607 (594~km), UK608 (602~km), FR606 (700~km).}
\end{figure}

The resulting distributions of coherence time for each international baseline are shown in Fig.~\ref{coherence}. There is a noticeable anticorrelation between coherence time and baseline length. The longest median coherence time, at just over 3 minutes, is seen on the DE609 baseline (Norderstedt), which at 227~km is close to being the shortest. We see median coherence times of between 80 and 110 seconds on the four longest baselines (580--600~km), and about 2 minutes for the other baselines. The coherence time of $\sim$80~s on the longest baseline (SE607 = Onsala) is likely to be typical of that seen on baselines to the new Polish stations.

At first sight this dependence of coherence time on baseline length is unexpected, because the LBCS observations should be in an ionospheric regime where a narrow field is observed by stations at separations of a few hundred kilometres, which see essentially uncorrelated regions of ionosphere (regime 2 of Lonsdale, 2005\nocite{lonsdale05,intema09}); investigations with the VLA suggest ionospheric patch sizes of a few tens of kilometres (Cohen \& R\"ottgering 2009)\nocite{cohen09}. We have verified that the dependence on baseline length is not a signal-to-noise artifact by considering high signal-to-noise sources, which have ``P'' solutions on all baselines, and further restricting the sample to sources for which our polynomial fits reproduce the phase solutions with low scatter. In both cases the correlation persists over the ensemble of sources, even though for a given source the fastest phase variations may sometimes occur on shorter baselines. This complex picture reflects the complicated nature of the ionosphere, which features short time- and spatial-scale disturbances together with larger travelling ionospheric disturbances (e.g. Intema et al. 2009; Fedorenko et al. 2013)\nocite{intema09,fedorenko13}. We defer a full discussion of the ionospheric constraints provided by LBCS to a future work, after completion of the full survey.

\subsection{Delay and delay stability}

Delays are seen on all baselines throughout the LBCS project. These are different for different baselines, and for different epochs, but generally range from 0--200~ns, with no discernible dependence on baseline length. Such delays are easily sufficient to decorrelate when averaging over bandwidths of a few MHz. In general, delay solutions are much noisier than phase solutions within the LBCS database, although delay solutions should vary on longer timescales than phase solutions. We find no evidence for significant ($>$1--2~ns) delay variations in 3 minutes by visual observations of a few sources where we have a sufficiently high signal--to--noise ratio to see delay variations confidently. In principle, LBCS ``P'' sources can be used for delay calibration, although in practice we suggest that two tests should be undertaken on such sources: first, the L-R difference in inferred delay on the delay calibration with the 3-minute solution should be close to zero, and secondly, the scatter on the 0.1-minute delay calibrations should be compatible with obtaining delay calibrations to a few nanoseconds on the 5--10 minute timescales over which the delay is unlikely to vary.

\subsection{Phase transfer across space}

As well as a uniquely large and homogeneous dataset, LBCS also provides numerous examples of sources that are very close to each other and are strong enough to have phase solutions derived for each source individually. The main disadvantage of LBCS, however, is the short integration time on each source. Normally, one uses a substantial fraction of the total bandwidth to calibrate LOFAR long-baseline observations, over the whole observing time; this allows one to image the phase calibrator source and to derive its structure by self-calibration. This cannot be done with only 3 minutes and 3~MHz of bandwidth. Moreover, the amplitudes of the LBCS data are not well calibrated, beyond a very basic a-priori calibration.

Useful information about phase transfer can, however, be derived. For each pair of LBCS sources with a separation of less than 2\degrees, phase solutions have been derived on short (12-second) timescales. These solutions are represented as $P_1(t)$ and $P_2(t)$ respectively; similar solutions $D_1(t)$ and $D_2(t)$ are derived for delays on each antenna. We have then simulated an artificial point source at the phase centre, with amplitude of 1~Jy, and applied to it the phase and delay corrections $P_1(t)-P_2(t)$ and $D_1(t)-D_2(t)$ respectively. This represents the corruption incurred by using source 1 to calibrate source 2, and contains two major terms: the atmospheric phase and delay difference between the two points, and the unknown structures of the two sources that affect the phase and delay differences.

The effect of these corruptions is to reduce the amplitude at the centre of the resulting image, after they have been applied. This measurement gives a pessimistic view of the actual error in phase transfer in an actual observation, because in practice the source structure of the calibrator can be derived, as mentioned above, by self-calibration. Nevertheless, we can use the amplitude-reduction as a quality indicator, and in particular it can be correlated with the separation of the sources, and with the degree of resolution of the calibrator on longer baselines. One would expect to see at least some correlation with transfer distance, assuming that the isoplanatic patch due to the ionosphere is of the order of a degree. If no correlation is seen, it is likely that the major effect is the unknown source structure.

\begin{figure}
\includegraphics[width=9cm]{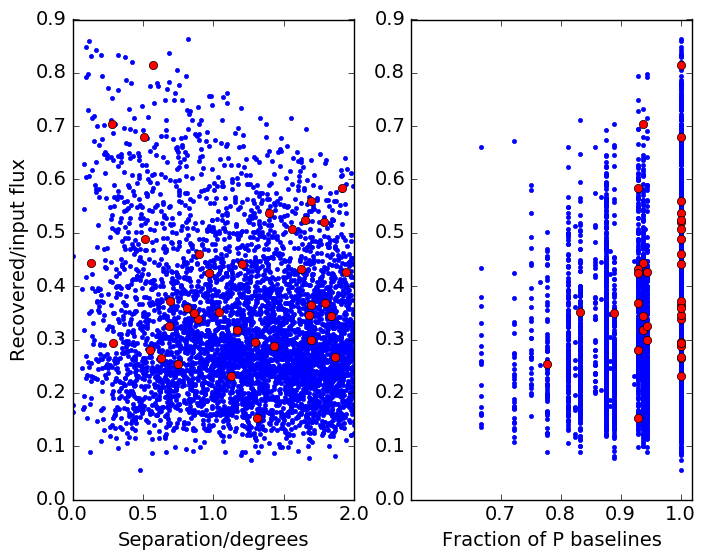}
\caption{Plots of the flux decrement, which is the maximum flux density observed for a source of unit amplitude that has been corrupted by phase and delay solution transfer from one of a pair of LBCS sources to the other (see text). Left panel: flux decrement against separation of the sources. Right panel: flux decrement against percentage of baselines to ST001 with clear source detections. In both cases the red circles indicate those cases where both LBCS sources in a given pair are also JVAS sources.}
\label{transfer}
\end{figure}

The results of the exercise are shown in Fig.~\ref{transfer}. The amplitude, after reduction from 1~Jy, is normally in the range 0.1--0.4~Jy, except at low separations where better results are achieved due to a tail of sources with unusually coherent phase transfer. The median reduction values change only slightly, from 0.32 at separations less than 0.5$^{\circ}$ to 0.28 at separations greater than 1.5$^{\circ}$.

The other effect on amplitude reduction is the source structure, and specifically the degree to which the source is resolved on the longer baselines. For each pair of sources we calculate the fraction of baselines to ST001 for which a ``P'' phase solution is obtained, and plot the amplitude reduction against this quantity also in Fig.~\ref{transfer}. It is evident that pairs of sources with a higher degree of resolution give generally worse phase transfer. For pairs where the fraction of ``P'' baselines is $\leq$0.8, the median amplitude, after reduction, is 0.25, and this rises to 0.3 where the fraction is $\geq$0.9; again, there is a tail of very good solution transfers where the source is well detected on all baselines.

We can also investigate the effect of source structure, again indirectly, by looking at the subset of LBCS sources that are also detected in the Jodrell Bank-VLA Astronometric Survey (JVAS, Patnaik et al. 1992\nocite{patnaik92}). JVAS sources are compact, flat-spectrum sources at GHz frequencies, and we would expect the assumption of a pointlike structure, which we have made in the fringe fitting, to be good in these cases. Fig.~\ref{transfer} shows that flux recovery is better for pairs of LBCS sources that are also JVAS sources compared to the overall distribution, and a K-S test shows that the distributions are different at better than 1\% significance, despite a relatively small ($\sim$0.07) difference in the medians of the samples. Inspection of the data for cases where the recovery is less good reveals differences in phase solutions in some cases, particularly for larger separations. In some cases where the separation is less, we see that the sources interfere with each other, so that the phase solution is degraded by beating in the visibility structure from the combined effect of the close sources.

In summary, it appears that under most circumstances, some level of phase transfer is possible on distances up to 2\degrees, but separations of less than 1\degrees\ allow, but do not guarantee, good phase transfer. We have shown that lack of knowledge of source structure is an impediment to the fringe fitting, and that longer observations of phase calibrators should allow better phase transfer than we have been able to achieve between pairs of LBCS sources. In some cases phase transfer fails not because of ignoring structure within a source, but because of ignoring the effects of nearby strong sources. Finally, it is likely that those pairs of LBCS sources that allow good phase transfer are those which lie within an isoplanatic patch as well as happening to approximate well to point sources. This implies that, on average, the isoplanatic patch for baselines of a few hundred kilometres is of the order of 1\degrees.

\section{Conclusion}

\subsection{Initial results from the LBCS survey}

The LBCS calibrator survey is an ongoing project to assess all bright sources, in the northern sky and with a relatively flat low-frequency radio spectrum, as possible long-baseline calibrators for LOFAR. About 15000 sources have been observed, and a further 15000 are planned for observing, to form the overall publicly available database. We have outlined procedures for making wide-field, fringe-rate and delay maps from the data. The database will contain plots of phase and delay solutions, summaries of coherence as a function of baseline, and eventually fringe-rate and delay maps.

The overall results of the survey are as follows:

\begin{itemize}
\item 49\% of sources examined are clearly detected, as indicated by the measures of scatter in phase solutions, on at least one baseline between international stations and the phased LOFAR core. Only 16\% of sources are clearly detected on all baselines.
\item In agreement with Mold\'on et al. (2015), we find a clear tendency for stronger sources, and sources with flat low-frequency spectral indices, to have a higher detection rate.
\item There is a strong inverse dependence of detection rate on baseline length, which accounts for nearly all of the difference between sources detected on one baseline and on all baselines. On the 200-km baselines, we find a density of more than 1 good calibrator per square degree nearly everywhere away from the Galactic plane, Cas~A and Cyg~A. On the longest baselines, this density drops by a factor of 2, and only sporadically reaches 1 per square degree.
\item The typical coherence time, as derived from our phase solutions, is between 1 and 3 minutes on the 200-km baselines, and about one minute on the longest international baselines. This result has been derived from a large number of observations, taken on different occasions and in widely different parts of the sky.
\item The ability to transfer phase solutions from one source to another depends on the distance between them, the amount of correlated flux density on each baseline, and the existence of a good model for the source. To the extent that snapshot observations allow us to disentangle these effects, it appears that the effective isoplanatic patch is usually about 1\degrees, with very good phase transfer typically being obtained only below this separation. It is likely that knowledge of the source model would allow good calibration below this separation, although we are unable to say for certain as the short observation time does not allow us to perform imaging or build source models.
\end{itemize}

\subsection{Calibration strategies for long-baseline LOFAR observations}

We conclude with some remarks about general calibration strategies for LOFAR international baselines.

A number of publications have already demonstrated the feasibility of imaging with the international baselines of LOFAR (e.g. Wucknitz 2010, Varenius et al. 2015, Moldon et al. in preparation)\nocite{varenius15a,wucknitz10a}. International-baseline calibration is challenging for three main reasons: 1) the  station-based propagation effects are rapidly variable with both time and direction; 2) the residual delays to be calibrated are both large (inhibiting the ability to average visibilities) and dispersive (making them more challenging to fit); and 3) there are few sources with sufficient correlated flux density at high spatial frequencies to act as calibrators, and very few have been identified. The LBCS project was devised to address the last problem, although it does not provide models for calibrators, as such a program would be prohibitively expensive in observing time. We note that eventually some equivalent result may emerge from large-scale sky surveys being conducted with LOFAR via the Surveys Key Science Project. In the meantime we may sometimes be able to use existing data for an initial model, preferably data that covers similar spatial frequencies at a similar observing frequency. For example, MERLIN array data at 408~MHz is a highly suitable starting model for the sources where such data exist (Stacey et al. 2016, in preparation).

We now describe a typical observing setup for a LOFAR international-baseline observation using LBCS calibrator sources.  Standard calibration steps should be taken to ensure the short baselines can be calibrated: at least one scan should be made on a bright ``Dutch array'' flux density calibrator such as 3C48, 3C147, 3C196, 3C295, or 3C380, and optionally a nearby gain calibrator can be observed periodically or continuously (alternatively, the target field can be imaged and used to derive gain corrections, but this requires additional processing).  These steps ensure that the tied station ST001 can be formed, and that the flux scale can be calibrated.

The target itself may be bright ($\gtrsim100$ mJy in a compact component), moderate ($10-100$ mJy in a compact component), or faint ($\lesssim$10 mJy in a compact component).  In the first instance, the target is probably itself a LBCS calibrator, and no external calibrator need be observed; all necessary bandwidth can be placed on the target, with any spare bandwidth optionally placed on nearby LBCS calibrator(s) as a check.   Otherwise, the nearest available good-quality LBCS calibrator should be selected, and available bandwidth is divided into two, with the target and calibrator observed in the same frequency range in the same time interval.

The first potential problem is the structure of the international-baseline calibrator.  In the most ideal case, it is known to be a point source, or a good model for it is available; in this case, the calibrator can be fringe-fitted to determine the delays\footnote{In practice, the delays typically vary less fast than the phase, and can probably be estimated on a longer timescale using a more distant source  \citep[e.g.,][]{varenius15a}}, then imaged and self-calibrated to optimize the model, and the amplitude, delay, and phase solutions can then be transferred to the target source.

In most cases, however, the calibrator structure is not known in advance. Tests with data obtained under the LOFAR Surveys Key Science Project suggest that it can be difficult to make good models for complicated sources in international-baseline observations, even 200-km ones, if the starting model is a point. In other cases, a point-source starting model has proved more successful (Varenius et al. 2016, in preparation). If the correct model is used, a hybrid mapping loop (self-calibration/imaging) converges quickly; otherwise the process does not converge and a map with high residuals is obtained. However, even an approximately correct model (e.g. a double source with the correct separation and amplitude ratio) is enough to begin the process of convergence. This can be done in one of two ways: using data at lower resolution \citep[e.g., the FIRST survey which has a resolution of 5$^{\prime\prime}$ at 20~cm;][]{becker95a}, or by fitting the closure phases by brute force using a grid-search through possible models. Software to do this does not currently exist, but is being developed.

Once delay and phase transfer from the calibrator has been accomplished, the phases on the target field will typically need to be further refined (unless the target itself was the calibrator, or the calibrator was particularly close to the target on the sky, within $\sim$10 arcminutes).  For a moderately bright target source, an imaging/self-calibration loop on the target will suffice, although it may be subject to the same starting model difficulties as the calibrator in the event that the calibrator--target separation is large ($\gtrsim1\degrees$).  For a faint target, it is instead necessary to locate a nearby background source (closer on the sky than the delay calibrator) to refine the transferred phases.  A fainter source can be used because self-calibration is more robust than fringe fitting, and because the coherence time has been increased allowing for longer solution intervals.  A minimum flux density of $\sim$10 mJy in a compact component is required, and the density of such sources is currently unknown.  However, all feasible candidates identified from lower-resolution imaging (or archival data) could be tested by repeatedly $uv$-shifting and averaging the target data set to candidate sources within $\sim$0.5$^{\circ}$ and imaging.  If this approach is taken, the visibility data cannot be heavily averaged prior to calibration, since this would limit the field of view to be too small.


As described above, for moderate to faint targets ($<$100 mJy on sub-arcsecond scales), we recommend by default allocating half the bandwidth to a separate station beam on the LBCS calibrator for international-baseline observations.  However, this may not be necessary in all cases.  The station beam at HBA frequencies has a diameter on the order of 2\degrees\ (frequency dependent); accordingly, if a LBCS calibrator is separated by less than 2\degrees\ from the target then the observations can be targeted in between the calibrator and target sources, and the visibility data can then be phase-shifted to the target and calibrator and averaged to create two datasets, each with the full bandwidth.  However, this increases the complications posed by the imperfectly modelled, frequency-dependent station beam to both the calibrator and the target source, and adds considerable processing time.  Accordingly, we recommend only using a single beam to cover both calibrator and target when the separation is quite small, $\lesssim$0.5\degrees, and complete frequency coverage is highly valued.


Finally, we remark on future calibration directions for LOFAR international-baseline observations.  To date, calibration of international-baseline data sub-divides the total observing bandwidth in order to approximate the changing dispersive delay with frequency as constant over a small bandwidth interval.  This is sub-optimal in terms of sensitivity, since not all of the bandwidth is utilised simultaneously.  Planned developments include a simultaneous fit for dispersive and non-dispersive delay over the whole observing bandwidth, but these require careful tuning to ensure that the global best solution is reached.  When available, this will improve calibration fidelity and (it is hoped) somewhat reduce the minimum flux density required for a delay calibrator.


\section*{Acknowledgements}
LOFAR, the Low Frequency Array designed and constructed by ASTRON, has facilities in several countries, that are owned  by various parties (each with their own funding sources), and that are collectively operated by the International LOFAR Telescope (ILT) foundation under a joint scientific policy. The research leading to these results has received funding from the European Commission Seventh Framework Programme (FP/2007-2013) under grant agreements No 283393 (RadioNet3), in the form of a travel grant to NJ to support data analysis, and 617199 (ALERT). AT acknowledges receipt of an STFC postdoctoral fellowship. ADK acknowledges financial support from the Australian Research Council Centre of Excellence for All-sky Astrophysics (CAASTRO), through project number CE110001020.

\bibliographystyle{apj}
\bibliography{lobos}

\bigskip

\bigskip

\appendix{\noindent \Large \bf Appendix A: The LBCS calibrator database}

We now describe the LBCS database, which is publicly available and is maintained at ASTRON at the webpage {\tt http://vo.astron.nl/}. The top-level database product is the calibrator list, the first page of which is shown in Table~\ref{caltable}. This includes a list of the pointing position, mostly taken from the WENSS survey and that have an accuracy of 1$^{\prime\prime}$--2$^{\prime\prime}$ for the strong sources considered here. It also includes the date and time of observation; some sources have been observed more than once, as previously discussed, to check reproducibility and data quality. The list also includes the classification described above (P, S, X, D or ``-'', where the latter symbol signifies that the station was not working) for each international station, ordered by station number from 601 to 609 (see Table 1). Finally, the table includes a number that is intended to give an idea of the quality of the observations. Because we do not have an independent measure of the observation quality, we use the percentage of sources detected in each pointing as an indicator, given that on average we expect to detect 30-40\% of sources on at least some baselines. Failure to detect more than 20\% of sources is unlikely, given our detection distribution, and fewer than 10\% almost certainly indicates that the observation failed for some reason.

The database is searchable via a standard cone search, where the user can specify a pointing direction and maximum radius. In addition to the flat-text list of calibrators, it contains the following items:

\setcounter{table}{0}
\renewcommand{\thetable}{A\arabic{table}}
\begin{table*}
\begin{verbatim}
L326144  00:00:00.990  68:10:03.000  2015-03-19  11:39:55  XXX-XXXXX  43
L269677  00:00:41.620  39:18:03.499  2015-03-05  12:38:02  XXXXXXXXX  53
L325856  00:00:41.620  39:18:03.499  2015-03-19  11:09:45  SXX-XXXXP  56
L269693  00:00:42.390  35:57:41.602  2015-03-05  12:38:02  PPPPPXXPP  53
L325860  00:00:42.390  35:57:41.602  2015-03-19  11:09:45  PXP-PPSPP  56
L410226  00:00:46.920  11:14:29.000  2015-11-09  19:30:06  XXXXXXXXX  30
L397819  00:00:49.470  32:55:47.701  2015-07-30  02:51:00  XSPSXPSSX  41
L269863  00:00:51.240  51:57:20.200  2015-03-05  12:44:02  PPPPPXPPP  53
L323730  00:00:51.240  51:57:20.200  2015-03-18  12:02:04  PPP-PXXPP  40
L269321  00:00:53.120  40:54:01.501  2015-03-05  12:26:02  PPPPPPPPP  63
L325788  00:00:53.120  40:54:01.501  2015-03-19  11:03:43  PPP-PPPPP  66
L410198  00:00:53.120  40:54:01.501  2015-11-09  19:36:08  PPPPPPPPP  70
L269649  00:00:54.520  38:02:44.999  2015-03-05  12:38:02  PXXXPXXXP  53
L325870  00:00:54.520  38:02:44.999  2015-03-19  11:09:45  PXP-PXXXP  56
L269313  00:01:01.520  41:49:29.201  2015-03-05  12:26:02  PPPPPXPPP  63
L325796  00:01:01.520  41:49:29.201  2015-03-19  11:03:43  PPP-PXPSP  66
L410206  00:01:01.520  41:49:29.201  2015-11-09  19:36:08  PPPPPSSSP  70
L410220  00:01:02.320  10:35:49.600  2015-11-09  19:30:06  XXXXXXXXX  30
\end{verbatim}
\caption{First few sources of the current LBCS catalogue. Columns are the LBCS observation number (equivalent to that maintained in the LOFAR Long-Term Archive); the right ascension of pointing position of the observation of the source; the declination of the pointing position; the observing date and time; the description of the source fringe detections to each of the LOFAR international stations (see text); and a quality flag for the observation in terms of the proportion of sources in a single observation that are detected. In general, below 20\% indicates a possibly bad observation and below 10\% indicates an observation that is likely to be bad.}
\label{caltable}
\end{table*}

\begin{itemize}
\item a png file of the form of Fig.~\ref{example}, i.e. a colour-scale image of the data phase $V(t,f)$ as a function of frequency on  baselines to ST001, and its Fourier transform, together with a plot of delay and phase solutions as a function of time;
\item a fringe-rate and delay map, in cases where detectable signal is observed. ``Detectable'' is defined as at least some phase solutions on 0.1-minute intervals being detected;
\item a data summary structure, that is readable with the {\tt numpy} routine {\tt numpy.load}. This is formatted as 
a Python dictionary and contains the phase and delay solutions, average $u$, $v$, and $w$ coordinates for the observation
and basic metadata such as observation date and time;
\item the {\sc aips} log file of the analysis of the {\sc fits} format data which produced the phase solutions.
\end{itemize}

\end{document}